\documentclass[CRPHYS,Unicode,manuscript]{cedram}
\usepackage{blindtext}
\usepackage{epsfig}
\usepackage{color}
\usepackage{amsmath}
\usepackage{amsfonts}
\usepackage{amssymb}
\usepackage{graphicx}%
\usepackage{url}
\usepackage{esvect}
\usepackage[normalem]{ulem}
\usepackage{xcolor}
\usepackage{xfrac}
\usepackage{dsfont}
\usepackage{bm}
\usepackage{array}
\newcolumntype{C}[1]{>{\centering\let\newline\\\arraybackslash\hspace{0pt}}m{#1}}
\usepackage{footnote}
\usepackage{isomath}
\DeclareMathAlphabet\mathbfcal{OMS}{cmsy}{b}{n}
\usepackage{esint} 
\usepackage{xcolor}
\usepackage[dvipsnames]{xcolor}

\hypersetup{colorlinks,citecolor=blue,filecolor=blue,linkcolor=blue,urlcolor=blue}

{
}

\title{Free Majorana Fermions with Superconducting Quantum Wires and a Magnetic Impurity}

\alttitle{Fermions de Majorana Libres avec des Fils Supraconducteurs et une Impuret\' e Magn\' etique}

\author{\firstname{Karyn} \lastname{Le Hur}\CDRorcid{0000-0002-3990-4782}\IsCorresp}
\address{CPHT, CNRS, Institut Polytechnique de Paris, Route de Saclay, 91120 Palaiseau, France}
\email[K. Le Hur]{karyn.le-hur@polytechnique.edu}

\begin{abstract} Through the two-channel Kondo model, I address a magnetic spin-1/2 impurity interacting with a bound state of spin origin at the edge in a Luther-Emery liquid showing a spin gap in the bulk. 
The system presents two zero-energy Majorana fermions of magnetic origin, one on the impurity site and one at the edge. I derive the wavefunction at the edge for the produced zero-energy Majorana
fermion bound state within the same formalism as topological interfaces. I present alternative versions of the quantum field theory revealing the specific pairing of the Majorana fermions in the bulk. I address local physical responses.
I develop the idea that this model can be realized with a magnetic impurity bridging the gap between two s-wave superconducting wires. I show the protection of the zero-energy Majorana fermions that will be referred to as free Majorana fermions.
  \end{abstract}
  
\begin{document}
  \maketitle

\section{Introduction}

The quest of Majorana fermions \cite{Majorana} attracts a lot of attention in physics \cite{Franz,MajoranazeroModes,Beenakker}, in particular, for their potential application in quantum information through topological systems \cite{Bert} related to non-Abelian anyons and fractional statistics \cite{MooreRead,Ivanov,Review,KitaevReview}. The protection from a topological state of matter such as the $p+ip$ superconductor in two dimensions \cite{ReadGreen} and the Kitaev p-wave one-dimensional (1D) 
superconducting wire \cite{Kitaev} refers specifically to the Majorana zero modes i.e. to Majorana fermions at zero energy topologically protected from the bulk. Analyzing bound state solutions in vortex cores of superconductors \cite{Caroli,KopninSalomaa,Volovik}, this precisely gives rise to a zero-energy Majorana mode per vortex for the $p+ip$ superconductor as a result of symmetries \cite{ReadGreen}. Current noise can detect the presence of two Majorana fermions belonging to the cores of different vortices at zero energy
\cite{BolechDemler}. Realizing a proximity effect between surface states of strong topological insulators and an s-wave superconductor \cite{FuKane} is also a way to produce such a Majorana fermion. Generalizations in graphene admit near zero modes \cite{DoronKaryn,GhaemiWilczek}. Various proposals do exist to realize a 1D p-wave superconductor \cite{Kouwenhoven,Refael,Maryland} including alternative versions with quantum spin models \cite{XMi,fractionalinfo,J1J2} and smaller systems of two dots \cite{Flensberg,Delft1,Delft} or two spins \cite{fractionalinfo}. The relation between Majorana fermions and topological aspects yet gives rise to fractional topological numbers for quantum spin models \cite{fractionalinfo,KLHReview,HH}. Zero-energy Majorana fermions also exist in the two-channel Kondo model \cite{NozieresBlandin,EmeryKivelson,ClarkeGiamarchiShraiman,SenguptaGeorges,Affleck,EntropyInfo} when coupling a magnetic impurity to two channels of conduction electrons. In that case, the two-channel Kondo model in a metal is unstable towards a channel anisotropy. In 1999, through a magnetic impurity on one edge of a ladder describing a d-wave superconductor \cite{KarynEPL}, I derived an {\it effective model} for a magnetic impurity interacting with the spin bound state of a Luther-Emery liquid \cite{LutherEmery,FabrizioGogolin} through the two-channel Kondo model. This article was motivated from the general study of bound state effects in superconductors with d-wave symmetry in a ladder geometry related to the physics of high-Tc superconductors \cite{KarynMaurice,LinBalentsFisher}. The same year, we also introduced a p-wave superconducting phase in those ladders \cite{UrsKaryn}. This is then the starting point of this article today.
One justification for this article is that the effective model is in fact general and can be thought of as a Majorana bound state in a 1D massive Dirac theory interacting with a localized Majorana fermion. I will analyse from different angles the quantum field theory aspects associated to this model showing the specific binding of the Majorana fermions of magnetic origin in the bulk. I formulate an analogy to topological interfaces and the Jackiw-Rebbi model \cite{JackiwRebbi,Shenbook,articlefluctuation} when solving the wavefunction of the zero-energy Majorana mode at the edge. I emphasize that zero-energy Majorana fermions in this article are of magnetic origin related to the spin sector of the Luther-Emery liquid whereas the charge sector forms the superfluid ground state \cite{LutherEmery,Giamarchi}. I show a correspondence between the local capacitance at the edge in the Kitaev p-wave superconducting wire and the magnetic susceptibility on the impurity site in this model. I present the application of these ideas to a magnetic impurity such as $Cu$ in between two s-wave superconducting wires such as $Al$ and $Nb$ in the weakly attractive limit. This study is motivated from recent thoughts on realizing the Kitaev p-wave superconductor with magnetic impurities on top of an s-wave superconductor \cite{StephanReview}. I address the protection through the spin gap in the bulk and the resonance with the impurity. 

The article is organized as follows. In Sec. \ref{Model}, I introduce the model. Since the Hilbert space on the impurity is a physical spin-1/2, this model also ensures the existence of a zero-energy Majorana fermion, referring to the {\it Majorana impurity} $a$, and the existence of a {\it zero-energy Majorana fermion within the bound state in the wire} $\gamma_b$. I derive the wave function associated to this Majorana fermion $\gamma_b$ through an analogy to topological interfaces.
I derive equivalent forms of the quantum field theory for the bulk and show the structure of the bonding of Majorana fermions $\gamma_a$ and $\gamma_b$ within the bulk. In Sec. \ref{correspondences}, I evaluate the local capacitance at the edge for the 
Kitaev p-wave superconducting wire and show a similar form as the magnetic susceptibility on the impurity for the present model. In Sec. \ref{wiremodel}, I address its specific realization with a magnetic impurity between two s-wave superconducting wires. In Appendix \ref{dictionary}, I derive useful mathematical correspondences for the magnetic impurity between two s-wave superconducting wires. In Appendix \ref{twoimpurities}, as a perspective, I address the situation of two magnetic impurities.

\section{Majorana Fermion in a Superconducting Wire with a Majorana Impurity}
\label{Model}

Then, we begin with the model through the Hamiltonian ${H}={H}_0+{H}_m+{H}_c$ \cite{KarynEPL}
\begin{eqnarray}
\label{model}
{H}_0 &=& - i\hbar v_F\int_{-L}^{L} dx c^{\dagger}(x)\partial_x c(x) \\ \nonumber
{H}_m &=& \int_{-L}^{+L} dx \left(-i\Delta c^{\dagger}(x) c(-x) sgn(x)\right) \\ \nonumber 
{H}_c &=& \lambda i(c(0)+c^{\dagger}(0))b.
\end{eqnarray}
A spin-1/2 impurity admits the following representation with two Majorana fermions $a=\frac{1}{\sqrt{2}}(d+d^{\dagger})=\sqrt{2}S_x$, 
$b=\frac{1}{\sqrt{2}i}(d^{\dagger}-d)=\sqrt{2}S_y$ and $S_z=i b a$ from the Jordan-Wigner transformation (for one site). Here, $a$ and $b$ are Majorana fermions such that $a^2=b^2=\frac{1}{2}$
and $\frac{1}{2}(S_x+i S_y)=d^{\dagger}$. I emphasize here that $a$ refers to a zero-energy Majorana fermion on the impurity site i.e. to the Majorana impurity.
The model ${H}_0+{H}_c$ is precisely the form of the {\it spin} Hamiltonian for the two-channel Kondo model of Nozi\` eres-Blandin in a metal at the Emery-Kivelson point \cite{NozieresBlandin,EmeryKivelson}.
Here, $H_0$ describes the Hamiltonian associated to spin degrees of freedom in the bulk and $H_c$ is the coupling with the impurity. 
The term $H_m$ then produces an energy gap in the energy spectrum for the {\it spin degrees of freedom} or equivalently for the fermion $c(x)$ related to the physics of the Luther-Emery model
 \cite{FabrizioGogolin} and we can assume $\Delta$ to be real. For simplicity, charge degrees of freedom are hidden and form the gapless superfluid motion of Cooper pairs in one dimension. We will re-instore them in the Appendix \ref{dictionary} when addressing
 the situation of an impurity in between two s-wave superconductors. We will verify e.g. that the gapless charge sector also supports the existence of a zero-energy bound state of spin origin in the Luther-Emery liquid.
 The symbol $H_m$ means that it also produces a mass to the Dirac fermions. When we take the limit $x\rightarrow 0$, i.e. going to an edge, the mass term $\Delta$ goes to zero as a result of the $sgn(x)$ function. 
 This will allow for the presence of zero-energy modes at the edge of the wire in the spin sector. If we introduce the Majorana fermions $\gamma_a=\frac{1}{\sqrt{2}}(c+c^{\dagger})$ and $\gamma_b=\frac{1}{\sqrt{2}i}(c-c^{\dagger})$
 in the bulk then the fermion $\gamma_b$ remains free close to the impurity. Therefore, the model in Eq. (\ref{model}) also describes a Majorana bound state $\gamma_a$ at the edge of the Luther-Emery liquid or in a 1D massive Dirac theory
 interacting with a local Majorana fermion $b$ on the impurity site such that these two Majorana fermions are paired together.
  
 \subsection{Majorana Fermion WaveFunction at the Edge}
 
 First, I shall derive the wavefunction associated to this zero-energy Majorana bound state $\gamma_b$ at the edge. I generalize the approach of Fabrizio-Gogolin
 for the Luther-Emery liquid \cite{FabrizioGogolin} in the presence of the impurity that will then allow for Majorana solutions for bound states at interfaces showing in the resolution a parallel with topological interfaces in the sense of 
 Jackiw-Rebbi \cite{JackiwRebbi,Shenbook,articlefluctuation}. We look for solutions in energy space, with $\epsilon$ the energy, of the form $c(x)=\sum_{\epsilon} c_{\epsilon} \chi_{\epsilon}(x)$ such that the Hamiltonian $H_0+H_m$ can be written as $\sum_{\epsilon} \epsilon c^{\dagger}_{\epsilon} c_{\epsilon}$. Here, $\chi_{\epsilon}(x)$ is assumed to be real.
 From the correspondence of Hamiltonians in {\it energy} and {\it real space}, we can then write down
\begin{eqnarray}
\label{one}
[H,c(x)] &=& -\sum_{\epsilon} \epsilon c_{\epsilon} \chi_{\epsilon}(x) -i \lambda b \sum_{\epsilon}\chi_{\epsilon}(0)\chi_{\epsilon}(x)\delta_{x0} \\ \nonumber
&=& i \hbar v_F \partial_x c(x) + i \Delta c(-x) sgn(x) -i \lambda b \delta_{x0}.
\end{eqnarray}
The last term of each line comes from the coupling to the magnetic impurity i.e. at $x=0$. The identification between these two terms leads to the normalization of the wavefunction $\sum_{\epsilon}\chi^2_{\epsilon}(0)=1$; we look for
a real wavefunction symmetrically localized around $x=0$. When $\lambda=0$, the equation for the zero-energy bound state is indeed very similar to the one in the Jackiw-Rebbi model presenting a protected bound state at an interface
with a potential difference in the Dirac equation \cite{Shenbook,articlefluctuation}. 
We also have
\begin{eqnarray}
\label{two}
[H,c^{\dagger}(x)] &=& \sum_{\epsilon} \epsilon c^{\dagger}_{\epsilon} \chi_{\epsilon}(x) -i \lambda b\delta_{x0} \\ \nonumber
&=& i \hbar v_F \partial_x c^{\dagger}(x) + i \Delta c^{\dagger}(-x) sgn(x) -i \lambda b \delta_{x0}.
\end{eqnarray}
Subtracting these two equations corresponds on the right-hand side to look for a zero-energy solution associated to the local operator $c(x)-c^{\dagger}(x)$ i.e. to the $\gamma_b(x)$ Majorana fermion. The information on the impurity simplifies
in this case. On the left-hand side, we add the energy of a hole $-\epsilon$ and subtract the energy of an electron $\epsilon$. Therefore,
\begin{equation}
-\epsilon(c_{\epsilon}+c^{\dagger}_{\epsilon})\chi_{\epsilon}(x) = i \hbar v_F \partial_x \chi_{\epsilon}(x)(c_{\epsilon}-c^{\dagger}_{\epsilon})+ i\Delta \chi_{\epsilon}(x)sgn(x)(c_{\epsilon}-c^{\dagger}_{\epsilon}).
\end{equation}
At zero energy $\epsilon=0$, this results in the equation 
\begin{equation}
(i\hbar v_F \partial_x \chi_{\epsilon=0}+ i \Delta \chi_{\epsilon=0}sgn(x))(c_{\epsilon=0}-c^{\dagger}_{\epsilon=0})=0,
\end{equation}
which admits the solution
\begin{equation}
\label{wavefunction}
\chi_{\epsilon=0}=\sqrt{\frac{\Delta}{\hbar v_F}} e^{-\frac{\Delta}{\hbar v_F} |x|}.
\end{equation}
The length scale $\xi=\frac{\hbar v_F}{\Delta}$ may be important $(\sim 0,1\mu m)$ if the superconducting (or in 1D spin) gap is of the order of $\sim 10$ Kelvins (compared to the electron bandwidth or Fermi energy $\sim 1eV$) 
i.e. assuming usual s-wave superconductors such as $Al$ or $Nb$. If $\Delta=0$, the latter would completely delocalize into the bulk with a probability $\sim 1/L$ to be present on each site. 
When $\Delta \neq 0$, this zero-energy Majorana solution is also more protected from perturbations such as anisotropies in the two channels representing the two wires; see Section \ref{wiremodel}.
If we sum these two equations there is an additional term $-2i\lambda b$ such that this does not reveal a free Majorana fermion $\gamma_a$. 

I have normalized the wavefunction on the whole space $]-L;L[$ with $L\rightarrow +\infty$. This implies that the Majorana bound state at zero energy delocalizes on the two sides of the impurity symmetrically.
As I show just below in Eq. (\ref{wireH0}), the model can also be presented for a wire of length $L$. 

\subsection{Alternative Forms of Quantum Field Theory}

It is important to emphasize that the Hamiltonian $H$ can equally describe the physics of a wire of length $L$. Indeed, we have the equivalent form for the Dirac Hamiltonian $H_0$ as
\begin{equation}
\label{wireH0}
H_0 = -i\hbar v_F\int_0^L dx c^{\dagger}(x)\partial_x c(x) +i\hbar v_F\int_0^L dx c^{\dagger}(-x)\partial_x c(-x),
\end{equation}
and the mass term $H_m$ can be re-written as
\begin{equation}
\label{wireHm}
H_m = \int_0^L dx\left(-i\Delta c^{\dagger}(x)c(-x) +i\Delta c^{\dagger}(-x)c(x)\right).
\end{equation}
To acquire more physical insight on the physical meaning of the model above, then we can write the model for a wire of length $L$ with {\it left} and {\it right} fermions 
such that $c(x)=c_R(x)$ and $c_L(x)=-c(-x)$, and at the edge $x=0$, $c_R(0)+c_L(0)=0$. The Hamiltonian can be reformulated as a quantum field theory $H=H_0+H_m+H_c$ in a wire of length $L$ with fermions moving in both directions:
\begin{eqnarray}
H_0 &=& -i \hbar v_F\int_0^L dx (c^{\dagger}_R(x) \partial_x c_R(x) - c^{\dagger}_L(x) \partial_x c_L(x)) \\ \nonumber
H_m &=& \int_0^L dx(i \Delta c^{\dagger}_R(x) c_L(x) +h.c.) \\ \nonumber
H_c &=& \lambda i (c_R(0)+c_R^{\dagger}(0))b.
\end{eqnarray}
The backscattering term $H_m$ describing the formation of the spin gap in the bulk is similar as in a band (Mott) insulator or Thirring massive model \cite{Giamarchi}, and the prefactor $i$ will have its importance when studying the wave-function of bound states or Majorana fermions solutions at an edge. The model $H_0+H_m$ can be refermionized introducing Fourier modes of right- and left- movers $c_{R,k}$ and $c_{L,k}$ respectively \cite{Giamarchi}. Then, we obtain the equivalent form for the bulk Hamiltonian
\begin{eqnarray}
H_0 &=& \sum_k (\hbar v_F k)(c^{\dagger}_{Rk} c_{Rk} - c^{\dagger}_{Lk} c_{Lk})\\ \nonumber
H_m &=& \sum_k \left(i\Delta c^{\dagger}_{Rk}c_{Lk}+h.c.\right).
\end{eqnarray}
This Hamiltonian can be written as a $2\times 2$ matrix in the spinor basis $(c_{Rk}, c_{Lk})$ and it is immediate to verify the form of the spectrum $\pm \sqrt{(\hbar v_F k)^2 +\Delta^2}$.
This model is general and describes the formation of an energy gap between valence and conduction bands.

Here, I give further physical aspects to this quantum field theory in terms of Majorana fermions. We can then introduce the Majorana fermions $\gamma_a=\frac{1}{\sqrt{2}}(c+c^{\dagger})$ and 
$\gamma_b=\frac{1}{\sqrt{2}i}(c-c^{\dagger})$ for left and right fermions such that
\begin{equation}
\frac{(-i\hbar v_F)}{2}\left(\gamma_a^p(x)\partial_x\gamma_a^p(x)+\gamma_b^p(x)\partial_x\gamma_b^p(x)\right)=\frac{(-i\hbar v_F)}{2}\left(c^{\dagger}_p(x)\partial_x c_p(x) + c_p(x)\partial_x c^{\dagger}_p(x)\right)
\end{equation}
and
\begin{equation}
i\Delta\left(\gamma_a^R(x)\gamma_a^L(x) + \gamma_b^R(x)\gamma_b^L(x)\right) = i\Delta\left(c^{\dagger}_R(x)c_L(x)-c^{\dagger}_L(x)c_R(x)\right).
\end{equation}
In this way, the Hamiltonian is transformed as
\begin{eqnarray}
\label{Majorana}
H &=& \sum_{p} \frac{(-i p \hbar v_F)}{2} \int_0^L dx \left(\gamma_a^p \partial_x \gamma_a^p + \gamma_b^p \partial_x \gamma_b^p\right)  \\ \nonumber
&+& \int_0^L  dx\ i\Delta \left(\gamma_a^R(x) \gamma_a^L(x) +  \gamma_b^R(x) \gamma_b^L(x)\right) \\ \nonumber
&+& \lambda i \sqrt{2} \gamma_a^R(0) b.
\end{eqnarray}
The sum on $p$ corresponds to the two directions of propagation $R,L$ or equivalently to $\pm 1$. 
Within these definitions, $i\gamma_a \gamma_b = c^{\dagger} c-\frac{1}{2}$. The Hamiltonian is also invariant under the transformation $\gamma_b\rightarrow -\gamma_b$.
At the impurity site, the boundary condition $c_L(0)+c_R(0)=0$ is also equivalent to $c_L^{\dagger}(0)+c_R^{\dagger}(0)=0$. Adding these equations, we obtain
a boundary equation formulated in terms of Majorana fermions e.g. 
\begin{equation}
\gamma_a^R(0)=-\gamma_a^L(0).
\end{equation}
 Subtracting these two equations, this results in 
 \begin{equation}
 \gamma_b^R(0)=-\gamma_b^L(0). 
 \end{equation}
 This is precisely the boundary condition imposed by the impurity. It provides a $\pi$-phase shift between left- and right Majorana fermions.
 
\begin{figure}[t]
\includegraphics[width=0.6\textwidth]{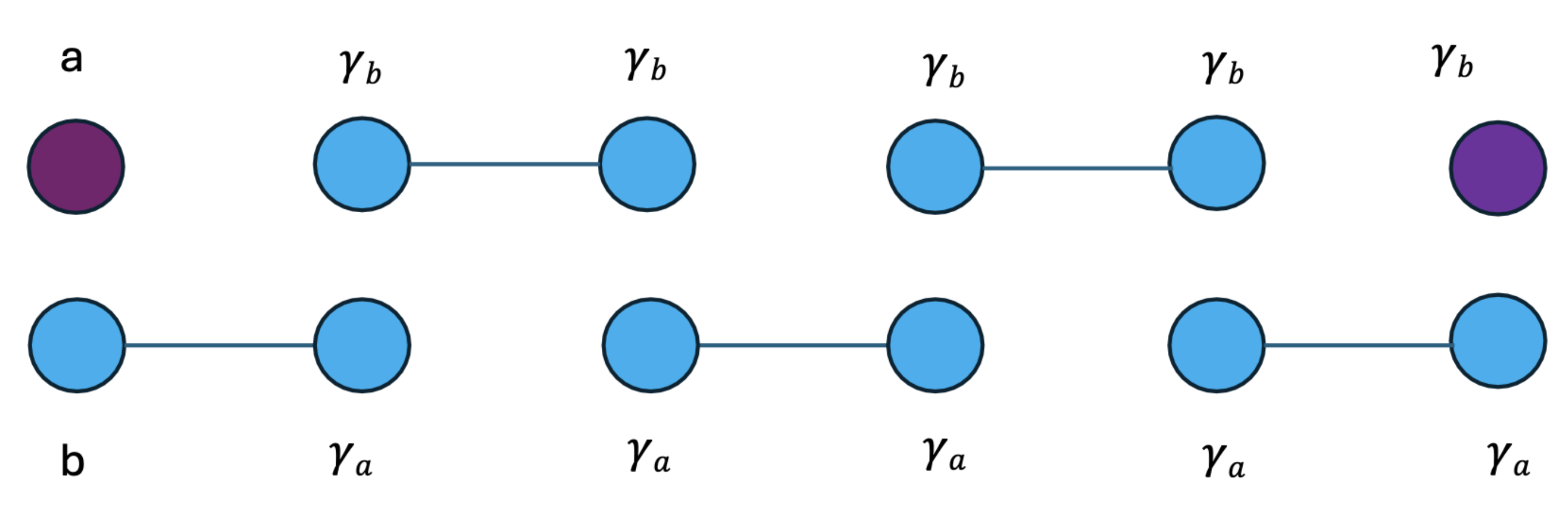} 
\vskip -0.3cm
\caption{Majorana fermions organization for the spin Hamiltonian $H=H_0+H_m+H_c$ in Eq. (\ref{Majorana}). The two zero-energy Majorana fermions, i.e. the free Majorana fermions, are represented in purple. The magnetic impurity is shown on the left through the Majorana fermions $a$ and $b$. Since the zero-energy solution for the wave-function associated to the  Majorana fermion $\gamma_b$ produces a typical length scale $\xi=\frac{\hbar v_F}{\Delta}$, we can place (draw) this particle at a certain distance from the Majorana fermion $a$. In Sec. \ref{wiremodel}, I show that the two Majorana fermions in purple are protected through the realization with a magnetic impurity producing a bridge between two s-wave superconducting quantum wires with singlet pairing.
In that case, the Majorana fermion $\gamma_b$ will be delocalized in a symmetric way between the two wires such that the characteristic length $\xi$ will develop on both sides of the impurity. The Majorana fermion $\gamma_b$
will resonate between the two wires.}
\label{MajoranaString}
\end{figure}

In the bulk, the effect of the term $\Delta$ is similar to produce a pairing between Majorana fermions with same flavor i.e. $a$ or $b$ on adjacent sites. Indeed, in the quantum field theory $\gamma_a^R \gamma_a^L=\gamma_a^R(x^-) \gamma_a^L(x^+)\rightarrow \gamma_a(i)\gamma_a(i+1)$. Similarly, the coupling between the Majorana fermions $\gamma_b$ on adjacent sites is important to produce a gap in the bulk in all the quantum fields or particles. These quantum field theories for Majorana fermions $\gamma_a$ and $\gamma_b$ are called {\it massive} and present an analogy with the 2D classical Ising model \cite{karyn1999,ItzyksonDrouffe}. This form of Hamiltonians is obtained for a mass term $i(c^{\dagger}_L c_R + h.c.)$ in the bulk and it can also be 
derived in a similar way with a pairing term
$c^{\dagger}_L c^{\dagger}_R+h.c.$ \cite{karyn1999}. Similarly as the p-wave Kitaev superconducting wire within the topological phase \cite{Kitaev}, this model can then be seen as a {\it string of paired Majorana fermions} \cite{LoicChristopheKaryn,FrederickLoicKaryn} with two (free) purple sites corresponding in the present case to a zero-energy Majorana fermion $a$ and to a zero-energy Majorana fermion $\gamma_b$. The terms $\Delta$ and $\lambda$ then produce the Majorana fermions structure in Fig. \ref{MajoranaString}. At the impurity site, the Majorana fermion $a$ is free in purple at zero energy. The fermion $\gamma_b$ close to the edge is also free; I draw it in purple at a distance place to emphasize that in this model the superconducting gap will produce a characteristic length scale $\xi=\frac{\hbar v_F}{\Delta}$ when solving the wave-function of the zero-energy solutions at the edge. The Majorana fermion $\gamma_b$ will have a probability one to be present in a region 
$x\in [-\xi;\xi]$ from the impurity or from the edge. 
It is then important to re-emphasize here that this corresponds to a magnetic analogue of the p-wave superconducting wire i.e. that $a$ and $\gamma_b$ are from the spin sector of the Luther-Emery Hamiltonian which develops a gap in the bulk
(see Sec. \ref{wiremodel}).

To build the magnetic analogy with the p-wave Kitaev superconductor \cite{Kitaev} from the bulk theory, we can introduce a particle-hole transformation such that 
$c(x)=c_R(x)$ and $\tilde{c}^{\dagger}(-x)=c(-x)=-c_L(x)=-\tilde{c}^{\dagger}_L(x)$. In this way, the Hamiltonian $H=H_0+H_m+H_c$ becomes
\begin{eqnarray}
\label{bulkpwave}
H_0 &=& i\hbar v_F \int_0^L  dx(\tilde{c}_L^{\dagger}(x)\partial_x \tilde{c}_L(x)) - i\hbar v_F \int_0^L dx c^{\dagger}_R(x)\partial_x c_R(x)) \\ \nonumber
H_m &=& - i \Delta \int_0^L dx(\tilde{c}_L(x) c_R(x) - c^{\dagger}_R(x) \tilde{c}_L^{\dagger}(x)) \\ \nonumber
H_c &=& i\lambda (c_R(0)+c_R^{\dagger}(0))b.
\end{eqnarray}
The boundary condition associated to the transformation on left particles is equivalent to fix $c_R+\tilde{c}^{\dagger}_L=0$ and $c_R^{\dagger}+\tilde{c}_L=0$. Summing these two equations, we identify the same form of boundary condition allowed by the impurity term $\gamma_a^R=-\gamma_a^L$ if we introduce the Majorana fermions from the left branch as $\gamma_a^L = \frac{1}{\sqrt{2}}(\tilde{c}_L+\tilde{c}_L^{\dagger})$ and $\gamma_b^L = \frac{1}{\sqrt{2}i}(\tilde{c}_L-\tilde{c}_L^{\dagger})$. The pairing term then becomes odd under parity similarly as for p-wave orbitals. The pairing term can then be seen as a continuum limit of the lattice term $i\Delta c^{\dagger}_i \tilde{c}^{\dagger}_{i+1}+h.c.$ in the Kitaev lattice Hamiltonian \cite{Kitaev}. In this way, this real-space representation of $H_m$, in addition to the drawing in Fig. \ref{MajoranaString}, favors an analogy with the p-wave Kitaev superconducting wire which also admits a topological or geometrical interpretation as a monopole in the Nambu reciprocal space \cite{SatoAndo,article2025}. 

\section{Local Physical Responses}
\label{correspondences}

I discuss here local physical correspondences between the magnetic susceptibility when applying a local magnetic field on the impurity site and the local capacitance in the p-wave superconducting wire (within the
topological phase e.g. at half-filling) when applying a local variation of the potential. We have recently shown in Ref. \cite{J1J2} that this correspondence is appropriate when probing topological quantum phase transitions e.g. 
the quantum phase transition in the $J_1-J_2$ spin model through the logarithmic edge spin susceptbility as in the two-channel Kondo model \cite{NozieresBlandin,EmeryKivelson,ClarkeGiamarchiShraiman,SenguptaGeorges},
corresponding then to the local capacitance at an edge in the Kitaev p-wave superconductor at the topological phase transition. I generalize the analysis to the topological phase
of the Kitaev wire, at half-filling, and show that the local capacitance at the edge is also very similar to the spin magnetic susceptibility in the present model when including a spin gap in the bulk.

To fix the definitions, we remind that at half-filling $(\mu=0)$, the Kitaev model reads \cite{Kitaev}
\begin{equation}
H_{pwave}=-t\sum_i (c^{\dagger}_i c_{i+1}+h.c.) + \sum_i (\Delta c^{\dagger}_i c^{\dagger}_{i+1}+h.c.).
\end{equation}
Introducing the Majorana fermions operators $\eta_{1i}=\frac{1}{\sqrt{2}}(c_i+c_i^{\dagger})$ and $\eta_{2i}=\frac{1}{\sqrt{2}i}(c_i^{\dagger}-c_i)$ such that $c^{\dagger}_i c_i -\frac{1}{2} = i \eta_2 \eta_1$ then
\begin{eqnarray}
\label{pwave}
H_{pwave} = \left(t+\Delta\right)\sum_i i \eta_{1i}\eta_{2i+1} + \left(-t+\Delta\right)\sum_i i \eta_{2i}\eta_{1i+1}.
\end{eqnarray}
For $t=\Delta$ and $\mu=0$, we verify the existence of a zero-energy Majorana fermion of the form $\eta_2(x=0)$ in the continuum limit. This is also equivalent to
$\eta_2(x=0)=\eta_2^R(x=0)$ and $\eta_2^L(0)+\eta_2^R(0)=0$ introducing the left and right Majorana fermions.

From the Bardeen-Cooper-Schrieffer (BCS) wavefunction defined on the half Brillouin zone $k\in [0;\pi]$, it is possible to evaluate the Green's function of the $\eta_1$ Majorana fermion at $x=0$, which results in
\begin{equation}
\langle \eta_1(0,\tau)\eta_1(0,0)\rangle = \frac{1}{2}\frac{1}{M}\sum_{k\in [-\pi;\pi]} e^{-\frac{E_k \tau}{\hbar}}.
\end{equation}
Here, $M$ is the number of sites or equivalently the length of the wire since we fixed the lattice spacing to {\it unity}. The Green's function is in imaginary time to evaluate the free energy correction in the presence of a potential variation.
This result can be explicitly verified from mathematical derivations; see e.g. Ref. \cite{J1J2}.
For $\mu=0$ the energy spectrum related to Bogoliubov quasiparticles on the wave-vector space $k\in [0;\pi]$ takes the form
\begin{equation}
E_k = \sqrt{(2t)^2\cos^2 k + (2\Delta)^2 \sin^2 k}.
\end{equation}
When $t=\Delta$, then from the bulk Hamiltonian, we obtain 
\begin{equation}
\langle \eta_1(0,\tau) \eta_1(0,0)\rangle = \frac{1}{2} e^{\frac{-E_k\tau}{\hbar}}=\frac{1}{2} e^{\frac{-2\Delta \tau}{\hbar}},
\end{equation}
which also agrees with the form of the Hamiltonian in real space (the ground state is reached fixing the parity operator $i\eta_1(0)\eta_2(1)$ to $-\frac{1}{2}$ and producing an excitation in the Majorana $\eta_1(x=0)$ sector requires to
flip $i\eta_1(0)\eta_2(1)$ to $+\frac{1}{2}$ with an associated gap $t+\Delta$. On a site, $c^{\dagger}_i c_i=0$ or $1$ implying that a parity operator of the form $i \eta_1 \eta_2$ admits eigenvalues $\pm \frac{1}{2}$.). 
For $t=\Delta$ and $\mu=0$, the Majorana fermion $\eta_2(x=0)$ is free at zero energy such that its Green's function takes the form
\begin{equation}
\langle \eta_2(0,\tau) \eta_2(0,0)\rangle = \frac{1}{2} sgn(\tau).
\end{equation}
Suppose we measure the local capacitance at site $x=0$ at the edge. This results in a variation of energy related to the Hamiltonian $\delta H=\delta \mu c^{\dagger}(x=0) c(x=0)=-\delta \mu i \eta_1(0)\eta_2(0)$. The free
energy correction takes the form
\begin{equation}
\Delta F = \int_0^{\beta} d\tau \langle \delta H(\tau) \delta H(0)\rangle = \frac{\hbar}{8\Delta}(\delta\mu)^2 \left(1-e^{\frac{-2\Delta \beta}{\hbar}}\right).
\end{equation}
The charge is defined as $\langle Q\rangle = \partial\frac{\langle\Delta F \rangle}{\partial \delta\mu}$ and the quantum capacitance at the edge $x=0$ for the Kitaev model within the topological phase at half filling is
\begin{equation}
\label{capacitance}
C = \lim_{\beta\rightarrow +\infty}  \frac{\partial^2 \Delta F}{\partial \delta\mu^2} = \frac{\hbar}{4\Delta}.
\end{equation}
This is a similar behavior as the {\it local magnetic susceptibility} (related to the impurity) that I found for the model of Eq. (\ref{model}) when adjusting the edge coupling with the impurity or the local Kondo
resonance $\Gamma=\frac{\lambda^2}{v_F}$ (for a unit lattice spacing) \cite{EmeryKivelson} with the superconducting energy gap $\Delta$ \cite{KarynEPL}. The charge response comes from short time scales. At site $x=0$, the two Majorana fermions 
can offer a charge response to the local deformation of the potential $\delta\mu$ at the edge similarly as if $\Delta=0$, such that $\Delta F\sim (\delta\mu)^2 \int_0^{\frac{\hbar}{2\Delta}} d\tau \frac{1}{4}$; this is identical as if $\eta_1(\tau)\eta_1(0)\eta_2(\tau)\eta_2(0)\sim \eta_1^2 \eta_2^2 =\frac{1}{4}$. 
 In comparison, for a resonant level model (for a free $d$-fermion), the capacitance would be similar to the Curie form for the magnetic susceptibility $\sim 1/T$ with $T$ the temperature.
 In the bulk, within the topological phase, since the Green's function of the Majorana fermions $\eta_1$ and $\eta_2$ acquire the same exponential decay
this results in a halved capacitance compared to the edge $x=0$. It is relevant to mention efforts in probing capacitance responses in superconducting wires \cite{capacitance}. 
The result for the capacitance(s) shows some resemblance with another bulk observable, i.e. the prefactor of the linear bipartite charge fluctuations or quantum Fisher information density, for the p-wave Kitaev superconducting wire \cite{LoicChristopheKarynFluctuations}.
When $t\neq \Delta$, the structure of Majorana fermions in Eq. (\ref{pwave}) is modified 
such that the characteristic binding energy for the Majorana fermion $\eta_1$ is $2t=2\Delta\rightarrow t+\Delta$. 
In that case, the upper bound in short time physics $\hbar/(2\Delta)$ turns into $\hbar/(\Delta+t)$ and $C=\frac{\hbar}{2(\Delta+t)}$. For a time scale $\sim \hbar/(t+\Delta)$, the effect of any local perturbation in $t-\Delta$ is not yet visible (in comparison, this would generate a longer time scale $\sim \hbar/(t-\Delta)$) such that the Green's function for the Majorana fermion remains identical, in accordance with a free Majorana fermion.

Coming back to the model of Eq. (\ref{model}) with the magnetic impurity, it is interesting to mention that in fact two limits {\it exist} i.e. when the local coupling $\lambda$ or specifically the Kondo resonance satisfies either
$\Gamma\gg 2\Delta$ or $\Gamma\ll 2\Delta$ \cite{KarynEPL}. In the second (latter) situation $\Gamma\ll 2\Delta$ (and also for $\Gamma\sim\Delta$), the local magnetic susceptibility shows a similar form as Eq. (\ref{capacitance}) with an energy scale $\sqrt{2\Gamma\Delta-\Gamma^2}$ \cite{KarynEPL}.  When $\Gamma\gg 2\Delta$, the Green's function of the Majorana fermion $b$ would acquire a similar form as for the two-channel Kondo model in a metal i.e. $G_b(\tau)=\frac{1}{2}\frac{1}{\pi\tilde{\Gamma}} \frac{\pi/\beta}{\sin\frac{\pi\tau}{\beta}}$ with $\tilde{\Gamma}=\sqrt{\Gamma^2-2\Gamma\Delta}$, whereas the Green's function of the Majorana fermion $a$ maintains a free form $G_a(\tau)=\frac{1}{2}sgn(\tau)$. In that case, the local magnetic susceptibility shows a logarithmic behavior $\sim \ln \frac{T}{\tilde{\Gamma}}$ reminiscent of the two-channel Kondo model \cite{NozieresBlandin,EmeryKivelson,ClarkeGiamarchiShraiman,SenguptaGeorges}, but cutoff at the energy scale $\tilde{\Gamma}$ due to the presence of $\Delta$.  I show in Eq. (\ref{wavefunction}) that the existence of the free Majorana mode $\gamma_b$ is in fact independent of the ratio $\Gamma/\Delta$, and therefore the pair of Majorana zero modes $a$ and $\gamma_b$ can yet occur below the energy scale $\tilde{\Gamma}$. When $\Gamma=2\Delta$ this is similar as if the spin-1/2 impurity remains unscreened.

In this way, the magnetic impurity can be seen as a physical sensor of the presence of Majorana fermions at the edge in the wire i.e. the magnetic impurity response to the local magnetic field identically probes the response to the Majorana fermions $\gamma_a(|x|<\xi)\leftrightarrow b$ and $\gamma_b(|x|<\xi)\leftrightarrow a$.

\section{Realization With s-wave Superconducting Wires In The Weakly Attractive Limit}
\label{wiremodel}

I address here the possible realization of the model in Eq. (\ref{model}) and of free Majorana fermions with two s-wave superconducting wires, with weak attractive intrinsic BCS interactions, meeting around $x=0$ where a magnetic impurity i.e. a spin-1/2 will be placed. The physical situation then implies the presence of two wires e.g. two edges present around the magnetic impurity. It is then important to understand what are the physical meanings of the fermion $c$ 
and of the Majorana fermions in this case. The important message will be that when including an attractive interaction of approximately same amplitude in the two wires it will be yet possible to mix the spin channels (degrees of freedom) in both wires
to write down the complete spin Hamiltonian with $H_m$ and $H_c$ in terms of the spin-polarized fermion $\psi_{sf}$ playing the same role as the fermion $c$ in Eq. (\ref{model}). 
For each wire, it is in fact possible to navigate from $]-\infty;0[$ or $]0;+\infty[$ onto $]-\infty;+\infty[$, where we suppose the wires very long. As in the two-channel Kondo model \cite{NozieresBlandin,EmeryKivelson,ClarkeGiamarchiShraiman,SenguptaGeorges}, the fermion $\psi_{sf}$ will be built from a symmetric superposition of fields in the two wires such that in Fig. \ref{MajoranaString}, the Majorana fermion $\gamma_b$ associated to the fermion $\psi_{sf}$ will have the same probability to reside on the two wires i.e. the Majorana fermion $\gamma_b$ will resonate around the impurity on the characteristic length scale $\xi$. I will address the stability of the zero-energy Majorana fermions towards perturbations (such as weak interaction asymmetry in the two wires, weak asymmetry in the coupling between each wire and the impurity and inter-wires tunneling effects around the impurity) emphasizing on the protection from the spin gap in the bulk and from the edge(s) as for a topological interface. I also emphasize that the present geometry with a magnetic impurity sandwiched between two s-wave superconducting wires presents an advantage compared to the situation where the impurity would be on one edge of the wires placed in parallel. In the latter situation, the inter-wires hopping term $t_{\perp}$ developing along the wires, that may be larger than the superconducting gap or spin gap in each wire, can modify the low-energy fixed point \cite{LutherEmerywire}. In Ref. \cite{KarynEPL}, the superconducting system referred to an impurity on one edge of a d-wave superconducting state in a Hubbard ladder \cite{KarynMaurice,LinBalentsFisher}. The model was motivated from the physics of 2D high-Tc superconductors
related to similar mathematical symmetry analyses \cite{Zhang,KarynMaurice}. Below, I develop the key steps for the generalization with a magnetic impurity in between two s-wave superconducting wires and I will address the protection of the zero-energy
Majorana fermions. For the sake of clarity, I
introduce the Luther-Emery liquid and the precise bosonization dictionary in Appendix \ref{dictionary} with all the correspondences for this specific geometry. In Appendix \ref{dictionary}, 
I also emphasize on why the charge sector supports the existence of a bound state of magnetic
origin in the Luther-Emery liquid formed with an attractive interaction in a quantum wire leading then to a dominant s-wave pairing term. It will also support the zero-energy Majorana fermions solutions in the present model.

The spin sector of the two s-wave superconducting wires, one present on $x\in ]-\infty;0[$ and one on $x\in ]0;+\infty[$, takes the form  \cite{Giamarchi,LutherEmerywire}
\begin{equation}
\label{spinmodes}
H_{swave} = H_{s} =  H_{os} + \frac{g}{(2\pi a)^2}\int_{-\infty}^0 dx \cos \sqrt{8}\phi_{1s}(x) + \frac{g}{(2\pi a)^2}\int_{0}^{+\infty} dx \cos \sqrt{8}\phi_{2s}(x).
\end{equation}
I re-instore the lattice spacing $a$ or short-distance cutoff (with the same letter as one of the two Majorana fermions related to the impurity). Here, $H_{os}$ represents a Luttinger type Hamiltonian for spin excitations and the cosine Sine Gordon terms represent the pairing terms associated to the two Luther-Emery superconducting wires, where $g<0$ for {\it attractive interactions} \cite{LutherEmerywire}. The presence of the cosine terms will favor a spin gap in each wire i.e. will favor $\phi_{is}(x)\sim 0$ for the ground state
in the bulk to minimize classically the energy. In Ref. \cite{KarynEPL}, we rather have $g>0$ associated to repulsive interactions yet leading to a d-wave superconducting instability. For this specific geometry, the term $g$ can be equally written in terms of the boson fields 
$\phi_s(x)=\frac{1}{\sqrt{2}}(\phi_{1s}(-x)+\phi_{2s}(x))$ and $\phi_{sf}(x)=\frac{1}{\sqrt{2}}(\phi_{1s}(-x)-\phi_{2s}(x))$
such that it takes the form 
\begin{equation}
+\frac{2g}{(2\pi a)^2}\int_0^{+\infty} dx\cos 2\phi_s(x) \cos 2\phi_{sf}(x). 
\end{equation}
The ground state will then correspond to pin or fix the phases $\phi_s(x)$ and $\phi_{sf}(x)$ such that it minimizes energy with 
$\cos 2\phi_s=+1$ and $\cos 2\phi_{sf}=+1$. A small difference of interactions in the two wires would produce an additional term $\frac{\delta g}{(2\pi a)^2} \int_0^{+\infty} dx \sin 2\phi_s(x) \sin 2\phi_{sf}(x)$. 
As long as the asymmetry verifies $|\delta g|\ll |g|$ then the pinning of the cosine potentials will keep (maintain) the sine potential terms e.g. $\sin 2\phi_s(x)$ to zero. It is precisely the symmetric situation between
the two wires that will favor a symmetric resonant coupling with the magnetic impurity. 

It is important to mention that the theory of each wire can also be re-written on $x\in ]-\infty;+\infty[$ in a symmetric way when developing the cosine terms in Eq. (\ref{spinmodes}) such that 
\begin{equation}
\label{spinwholespace}
H_s=H_{os}+\frac{g}{2(2\pi a)^2}\sum_{i=1,2}\int_{-\infty}^{+\infty} dx e^{i\sqrt{8}\phi_{is}(x)}.
\end{equation}
The coupling with the impurity is usually introduced through the field $\Phi_{sf}$ at $x=0$ \cite{EmeryKivelson,ClarkeGiamarchiShraiman,SenguptaGeorges}, as defined in Appendix \ref{dictionary}. Therefore, introducing the variables
$\Phi_s(x)$ and $\Phi_{sf}(x)$ in Appendix \ref{dictionary}, the spin Hamiltonian can be equivalently written as
\begin{equation}
H_s=H_{os}-\tilde{g}\int_{-\infty}^{+\infty} dx \cos(\Phi_s(x)-\Phi_s(-x)) e^{-i(\Phi_{sf}(x)-\Phi_{sf}(-x))},
\end{equation}
where $\tilde{g}=\frac{|g|}{(2\pi a)^2}$. I have implicitly assumed that the Hamiltonian is hermitian to reach this form.
It is then useful to apply a self-consistent approach similar to the BCS theory. To identify a similar form as the term $H_m$ in Eq. (\ref{model}), the last step will be a refermionization of the variables as
\begin{eqnarray}
i\psi^{\dagger}_s(x) \psi_s(-x) \hbox{sgn}(x) &=& e^{-i(\Phi_{s}(x)-\Phi_{s}(-x))} \\ \nonumber
i\psi^{\dagger}_{sf}(x) \psi_{sf}(-x) \hbox{sgn}(x) &=& e^{-i(\Phi_{sf}(x)-\Phi_{sf}(-x))}.
\end{eqnarray}
The factor $i$ is important because it encodes the Campbell-Baker-Hausdorff formula $e^{A+B}=e^A e^B e^{-\frac{1}{2}[A,B]}$ with $e^{-\frac{1}{2}[A,B]}=e^{[\Phi_{is}(x),\Phi_{is}(-x)]}$ where $[\Phi_{is}(x),\Phi_{is}(-x)]=-i\pi\hbox{sgn}(x)$.
The fermion $\psi_s$ will not interact with the impurity such that we can evaluate $i\langle \psi^{\dagger}_s(x)\psi_s(-x)\hbox{sgn}(x)\rangle=c$ in a self-consistent way when invoking renormalization group ideas. For weak attractive interactions,
the term $\tilde{g}$ is relevant at the energy scale
\begin{equation}
k_B T \sim \Delta = \Lambda e^{-\frac{\pi v_F}{|\tilde{g}|}}.
\end{equation}
Here, $\Lambda$ is a high-energy cutoff of the order of the bandwidth. This energy scale traduces the formation of a spin gap. This is then equivalent to
\begin{equation}
|\tilde{g}|c=\Delta=\Lambda e^{-\frac{\pi v_F}{|\tilde{g}|}}.
\end{equation}
Then, the term $g$ or $\tilde{g}$ is then equivalent to $H_m$ in Eq. (\ref{model}) for the fermion $\psi_{sf}$:
\begin{equation}
H_m = -i\Delta\int_{-\infty}^{+\infty} dx \psi^{\dagger}_{sf}(x) \psi_{sf}(-x)\hbox{sgn}(x).
\end{equation}

From the correspondence of operators at $x=0$ in Appendix \ref{dictionary}, the coupling with the impurity around $x=0$ takes the same form as in the two-channel Kondo model at the Emery-Kivelson line in a metal with two conduction channels \cite{EmeryKivelson,ClarkeGiamarchiShraiman,SenguptaGeorges}. Assuming a strong longitudinal coupling with the impurity $J_z\sim 2\pi v_F$, with $v_F$ the Fermi velocity, is first justified physically if the attractive force in the superconducting wires is small. In this sense, the coupling with the impurity goes first to strong couplings \cite{NozieresBlandin}. Therefore, we obtain an impurity Hamiltonian of the form
\begin{equation}
H_c = (2\pi v_F - J_z) \psi_s^{\dagger}(0)\psi_s(0)(i b a) + \frac{J_{\perp}}{\sqrt{\pi a}}(\psi_{sf}(0)+\psi_{sf}^{\dagger}(0))(ib).
\end{equation}
Here, $J_{\perp}$ corresponds then to the transverse coupling with the impurity. In the presence of the spin gap, since the ground state will fix the phase $\Phi_s(0)$ (see discussion in Appendix \ref{dictionary})
it naturally implies that in fact $\psi^{\dagger}_s(0)\psi_s(0)\propto \partial_x\phi_s(0)\rightarrow 0$.
This is then important because if we include any realistic small deviation for the longitudinal coupling $J_z$ from $2\pi v_F$ then this implies that the first term yet tends to zero, thanks to the spin gap in the bulk stabilizing 
the form of $H_c$ as in Eq. (\ref{model})
\begin{equation}
H_c = \lambda i (\psi_{sf}(0)+\psi_{sf}^{\dagger}(0))b
\label{psiHc}
\end{equation}
with $\lambda = \frac{J_{\perp}}{\sqrt{\pi a}}$ such that the Kondo energy scale introduced in Sec. \ref{correspondences} reads $\Gamma=\frac{J_{\perp}^2}{\pi v_F a}$.
In Appendix \ref{dictionary}, I emphasize that the ground state in the bulk will favor $\Phi_{sf}\rightarrow 0$ such that it will stabilize this form of $H_c$ at the edge towards a channel anisotropy that usually results
in a $\sin \Phi_{sf}(0)$ term. In Appendix \ref{dictionary}, I also emphasize on the role of the spin gap in the bulk of each wire to protect the two-channel Kondo fixed point from additional inter-wires tunneling effects. 

It is important to underline that probing directly the free Majorana fermion $\gamma_b$ associated to $\psi_{sf}-\psi_{sf}^{\dagger}$ delocalized around the two wires is not as easy. However, as mentioned in the preceding Section,
when probing the local magnetic susceptibility on the impurity this will reveal the presence of a zero-energy Majorana fermion $a$ which also implies the presence of another zero-mode Majorana fermion in the wire(s) $\gamma_b$. 
Indeed, it is important to emphasize that 
from general Hilbert space structure, the Majorana fermions appear in pairs: e.g. $S_z=i b a$ and the density $\psi_{sf}^{\dagger}\psi_{sf}$ or $c^{\dagger}c$ in Section \ref{Model} also corresponds to a term $i\gamma_a \gamma_b$.
The Majorana fermion $\gamma_a$ is bound to the impurity or to the Majorana fermion $b$ through the term in $J_{\perp}$ implying then a free Majorana fermion $\gamma_b$ around the impurity. The two Majorana fermions $a$ and $\gamma_b$
remain at zero energy.

\section{Conclusion}
\label{results}

To summarize, I have presented several aspects of a 1D quantum field theory with a zero-energy Majorana bound state at an edge interacting with a localized Majorana fermion. This
model engenders a structure of paired Majorana fermions in the bulk supporting two stable zero-energy Majorana fermions, one on the impurity and one at the edge of the wire(s). 
I have shown how the local magnetic susceptibility acting on the impurity presents analogies with the capacitance measure at the edge of the p-wave superconducting Kitaev wire within the topological phase. 
A natural realization of this model is with a magnetic spin-1/2 impurity bridging the gap between two s-wave superconducting wires i.e. two Luther-Emery liquids. 
This work then emphasizes on the possible local engineering of zero-energy Majorana fermions through s-wave superconducting wires within the weakly attractive regime,
i.e. 1D massive quantum field theories, from the interplay between the two-channel Kondo model and the model of topological interface of Jackiw and Rebbi. 
From the recent interest on realizing Majorana zero modes with an array of magnetic impurities on top of an s-wave superconductor, I hope that this work also opens the path to realize stable zero-energy Majorana fermions
with one magnetic impurity. With two impurities coupled to 1D s-wave superconducting electrodes, it is then possible to engineer a delocalized pair of zero-energy Majorana fermions which may find applications in quantum information; see Appendix \Ref{twoimpurities} as a perspective. 
\\

The article of Ref. \cite{KarynEPL}, that was initiated in my PhD thesis at LPS Orsay and published at ETH Z\"urich, was dedicated to my father Joel. I also dedicate this article today to my mother Evelyne and to my family.

\appendix

\section{Dictionary for the Luther-Emery liquid and Boson Fields}
\label{dictionary}

The Luther-Emery liquid \cite{LutherEmery} is formed through an attractive interaction of the form $g\psi^{\dagger}_{iR\uparrow}\psi_{iL\uparrow}\psi^{\dagger}_{iL\downarrow}\psi_{iR\downarrow}$. The electron operator
with spin polarization $\alpha=\uparrow,\downarrow$ in the wire $i$ reads \cite{Giamarchi}
\begin{equation}
\psi_{ip\alpha}=\frac{\kappa_{i\alpha}}{\sqrt{2\pi a}} e^{i(p\phi_{i\alpha}(x)+\theta_{i\alpha}(x))},
\label{electronoperator1}
\end{equation}
with $p=\pm$ for right- and left fermions respectively (associated to positive and negative momenta (group velocities) in the band structure). The Klein factors ensure the anti-commutation relations between electrons of different spin species such that
$\kappa_{i\uparrow}\kappa_{i\downarrow}=\pm i$. For $g<0$, this usually favors the s-wave channel such that the ground state associated to Eq. (\ref{spinmodes}) shows $\phi_{is}=0$ for the spin sector showing the spin gap. This implicitly
assumes that we introduce the {\it charge} and {\it spin} modes for each wire $\phi_{ic,is}=\frac{1}{\sqrt{2}}(\phi_{i\uparrow}\pm \phi_{i\downarrow})$ and $\theta_{ic,is}=\frac{1}{\sqrt{2}}(\theta_{i\uparrow}\pm \theta_{i\downarrow})$. 
The singlet s-wave channel then reads
\begin{equation}
\psi^{\dagger}_{i R\uparrow}(x) \psi^{\dagger}_{i L\downarrow}(x) - \psi^{\dagger}_{i R\downarrow}(x) \psi^{\dagger}_{ i L \uparrow}(x) \propto e^{-\sqrt{2}\theta_{ic}}\cos(\sqrt{2}\phi_{i s}).
\end{equation}
The charge sector shows a quasi-long range superfluid order. At the edge of each wire, we should satisfy $\psi_{i L \alpha}=-\psi_{i R \alpha}$, as a Friedel phase from the presence of the edge, which leads to $e^{\pm i \sqrt{2}\phi_{i c}}=-1$
or $\phi_{ic}(0)=\frac{\pi}{\sqrt{2}}$. Since the charge mode $\phi_{ic}$ will be pinned (fixed) at $x=0$ this implies that the dual operator $e^{-i\sqrt{2}\theta_{ic}}$ fluctuates very strongly and average to zero, such that it will be possible to
identify zero-energy solutions at the edge from the spin sector only. 

Then, I introduce the precise vocabulary in the present situation in comparison also the two-channel Kondo effect in a metal at the Emery-Kivelson point \cite{EmeryKivelson}. The spin modes $\phi_{1s}(x)$ and $\phi_{2s}(x)$ of each wire
belong respectively to the domains $x\in ]-\infty;0[$ and $x\in ]0;+\infty[$. It is first useful to mix the spin modes as
\begin{eqnarray}
\label{equation1}
\phi_s(x) &=& \frac{1}{\sqrt{2}}(\phi_{1s}(-x)+\phi_{2s}(x)) \\ \nonumber
\phi_{sf}(x) &=& \frac{1}{\sqrt{2}}(\phi_{1s}(-x)-\phi_{2s}(x)).
\end{eqnarray}
Here, I implicitly assume that $x\geq 0$. This choice of representation takes into account the physical space of each wire. However, since we equivalently introduce the spin Hamiltonian of each wire on the whole axis $]-\infty;+\infty[$ in
Eq. (\ref{spinwholespace}), we would find equivalent results introducing the same spin modes mixing as in the two-channel Kondo effect in a metal \cite{EmeryKivelson}
\begin{eqnarray}
\label{equationmodes}
\phi_s(x) &=& \frac{1}{\sqrt{2}}(\phi_{1s}(x)+\phi_{2s}(x)) \\ \nonumber
\phi_{sf}(x) &=& \frac{1}{\sqrt{2}}(\phi_{1s}(x)-\phi_{2s}(x)).
\end{eqnarray}

The coupling with the impurity occurs at $x=0$ where Eqs. (\ref{equation1}) and (\ref{equationmodes}) are equivalent. Another interesting aspect of the quantum field theory associated to the two-channel Kondo effect is that we can rephrase the fields $\phi_s(x)$ and its dual analogue $\theta_s(x)$ as a field $\Phi_s(x)$ and similarly for the correspondence between $\phi_{sf}$, $\theta_{sf}$ and $\Phi_{sf}$ \cite{EmeryKivelson,ClarkeGiamarchiShraiman,SenguptaGeorges}. For this purpose, related to the spin Hamiltonian in Eq. (\ref{spinwholespace}) written on the domain $]-\infty;+\infty[$ for each wire, we re-write the electron operator in Eq. (\ref{electronoperator1})
\begin{equation}
\psi_{i p\alpha} = \frac{\kappa_{i\alpha}}{\sqrt{2\pi a}} e^{i(p\frac{1}{2}(\Phi_{i\alpha}(x)-\Phi_{i\alpha}(-x))+\frac{1}{2}(\Phi_{i\alpha}(x)+\Phi_{i\alpha}(-x))}
\label{electronoperator2}
\end{equation}
with again $p=\pm$ for right and left-moving fermions respectively. This leads to the identifications 
\begin{equation}
\psi_{i R\alpha}(x) = \frac{\kappa_{i\alpha}}{\sqrt{2\pi a}} e^{i \Phi_{i\alpha}(x)}
\end{equation}
and
\begin{equation}
\psi_{i L\alpha}(x) = \frac{\kappa_{i\alpha}}{\sqrt{2\pi a}} e^{i \Phi_{i\alpha}(-x)}
\end{equation}
such that $\hbox{lim}_{x\rightarrow 0} (\Phi_{i\alpha}(x) - \Phi_{i\alpha}(-x))=\pi$ to satisfy that  $\psi_{i L \alpha}=-\psi_{i R \alpha}$.
This also implies that the coupling with the impurity at $x=0$ can be equally written in terms of right or left fermions within a wire. 
Comparing the formulae (\ref{electronoperator1}) and (\ref{electronoperator2}), then we reach the identifications
\begin{eqnarray}
2\phi_{i\alpha}(x) &=& \Phi_{i\alpha}(x)-\Phi_{i\alpha}(-x) \\ \nonumber
2\theta_{i\alpha}(x) &=& \Phi_{i\alpha}(x) + \Phi_{i\alpha}(-x). 
\end{eqnarray}
Introducing the spin channels similarly as $\phi_s$ and $\phi_{sf}$
\begin{eqnarray}
\Phi_{s}(x) &=& \frac{1}{\sqrt{2}}(\Phi_{1s}(-x)+\Phi_{2s}(x)) \\ \nonumber
\Phi_{sf}(x) &=& \frac{1}{\sqrt{2}}(\Phi_{1s}(-x)-\Phi_{2s}(x)),
\end{eqnarray}
then from Eqs. (\ref{equation1}), we identify the precise correspondence
\begin{eqnarray}
2\phi_s(x) &=& (\Phi_s(x) - \Phi_s(-x)) \\ \nonumber
2 \phi_{sf}(x) &=&  (\Phi_{sf}(x) - \Phi_{sf}(-x)).
\label{correspondencesequations}
\end{eqnarray}
The spin sector of the Luther-Emery liquid favors $\phi_{is}=0$ for the ground state, or equivalently $\phi_s(x)=0$ and $\phi_{sf}(x)=0$ symmetrically, which means that $\Phi_s(x) - \Phi_s(-x)=0$ and that 
$\Phi_{sf}(x) - \Phi_{sf}(-x)=0$. If we invoke the parity symmetry because the two wires are now delocalized on the whole space $]-\infty;+\infty[$ then this corresponds to invert the role between $\Phi_s(x)$ and $\Phi_{s}(-x)$
and similarly between $\Phi_{sf}(x)$ and $\Phi_{sf}(-x)$. This implies that the ground state will then fix $(\Phi_{sf}(x) - \Phi_{sf}(-x))=(\Phi_{sf}(-x) - \Phi_{sf}(x))$ implying that $\Phi_s(x)=\Phi_{sf}(x)=0$ (similarly as $\phi_s$ and $\phi_{sf}$). 
The equation $x_1-x_2=-x_1+x_2$ is more generally satisfied when $x_i=-x_i$ implying then $x_i=0$. This also agrees with $\Phi_s(x) - \Phi_s(-x)=0$.
In the presence of the spin gap in the bulk, the coupling with the impurity will then be through the term $\cos\Phi_{sf}(0)$, referring to $\psi_{sf}^{\dagger}(0)+\psi_{sf}(0)$ in Eq. (\ref{psiHc}), stabilizing the two-channel Kondo fixed point.

It is important to emphasize that in the present geometry, the two-channel Kondo effect is also protected from additional inter-wires tunneling effects. 
Indeed,  if we analyze the terms coupling the two wires through the impurity, taking into account that 
both $\phi_{ic}$ and $\phi_{is}$ are 'pinned', this results in the identifications $\psi^{\dagger}_{1R\uparrow}\psi_{2R\downarrow} \propto e^{-\frac{i}{\sqrt{2}}(\theta_{1c}+\theta_{1s}-\theta_{2c}+\theta_{2s})}$
with $\theta_{ic,is}=\frac{1}{\sqrt{2}}(\theta_{i\uparrow}\pm \theta_{i\downarrow})$. Since the Luttinger parameters in the charge and spin sectors, associated to the Hamiltonians $H_o$ for each sector, satisfy $\frac{1}{2}\left(\frac{1}{K_c}+\frac{1}{K_s}\right)>1$ (this is due to $K_{s}<1$ as a result of the attractive interaction \cite{LutherEmerywire}) these inter-wires terms will not flow to strong couplings within the renormalization group method \cite{Giamarchi}, and similarly for the analogues of the Kane-Fisher terms \cite{KaneFisher} with pure charge transfer. Similarly as the situation with a strong repulsive interaction \cite{FabrizioGogolin} we find that the renormalization of the attractive interaction $g$ to strong coupling \cite{LutherEmerywire}
also leads to the stability of the two-channel Kondo effect.

\section{System with two magnetic impurities}
\label{twoimpurities}

Suppose we generalize this result and introduce another magnetic impurity $\boldsymbol\tau$ around the impurity $\mathbf{S}$. 
Through Jordan-Wigner transformation, the two spins-1/2 should satisfy commutation relations such that $\tau^+ = f^{\dagger} e^{i \pi d^{\dagger} d}$ and $\tau_z=f^{\dagger} f-\frac{1}{2}=i n m$ where $m$ and $n$ are two Majorana fermions
such that $m=\frac{1}{\sqrt{2}}(f+f^{\dagger})$ and $n=\frac{1}{\sqrt{2}i}(f^{\dagger}-f)$. 

In the low-energy limit, i.e. for temperatures smaller than $\Delta$ and than the Kondo energy scale $\Gamma=\frac{J_{\perp}^2}{\pi v_F a}$, then the direct Ising interaction
$S_z \tau_z = (d^{\dagger} d-\frac{1}{2})(f^{\dagger} f-\frac{1}{2})=(i ab)(i mn)$ is suppressed because the Majorana fermion $b$ is bound to the wires. The transverse spin interaction produces a term in $S^{+}\tau^-+h.c. = d^{\dagger}(1-2d^{\dagger} d)f+h.c.
= d^{\dagger} f+h.c. = \frac{1}{2}(a+i b)(m-in)+\frac{1}{2}(m+in)(a-ib)\rightarrow i na$. This interaction then will fix the parity operator between the two Majorana fermions $n$ and $a$ i.e. $\langle i na\rangle 
=-1$ for an antiferromagnetic interaction. Our goal is e.g. to realize a (protected) non-local spin-1/2 from the zero-energy Majorana fermion $a$ and from the impurity $\boldsymbol\tau$. To achieve this goal, we can then suppose that the second impurity also interacts with the same wires i.e. the two impurities are in parallel. This impurity can also realize a two-channel Kondo effect. To address this physics, this is similar as if the spin $\mathbf{S}$ or fermion $d$ would not be present such that we can simply write down, similarly as above, $\tau^+ = f^{\dagger}$ and $\tau_z=f^{\dagger} f-\frac{1}{2}=i n m$, $\psi_{sf}(x)\sim \frac{e^{i\pi f^{\dagger} f}}{\sqrt{2\pi a}} e^{i \Phi_{sf}(x)}$. In that case, the Majorana fermion $n$ will be bound to the wires such that $\langle i na\rangle=0$
and $\langle S^{\dagger} \tau^-\rangle=0$.

Now, we apply a local magnetic field on the impurities along the vector ${\bf e}_x+{\bf e}_y$. In the case of an AC magnetic field or in the case of an electromagnetic wave, it may show a useful oscillatory term $h=h_0 e^{-i \omega t}$. We address perturbations of the form
\begin{equation}
\delta H = -h (S_x+S_y) - h (\tau_x+\tau_y) = - \frac{h}{\sqrt{2}}(a+b)-\frac{h}{\sqrt{2}}(1-2d^{\dagger}d)(m+n).
\end{equation}
Due to the coupling $H_c$ this results in $\langle \cos \Phi_{sf} S_x\rangle=\langle (\psi_{sf}+\psi_{sf}^{\dagger})(ib)\rangle \neq 0$, which implies that any small magnetic field along $x$ direction will not modify the ground state i.e.
the $x$ component of the spin is already screened by the coupling to modes in the wire such that $\langle a\rangle=0$. We also have that 
$\langle S_y\rangle=0=\langle b\rangle$. This conclusion can also be justified as follows: the two impurities $\boldsymbol\tau$ and $\mathbf{S}$ play a symmetric role. 
Due to the commutation relations between spins and the Jordan-Wigner string $e^{i\pi d^{\dagger} d}=(1-2 d^{\dagger} d)$ in the second term since $\langle S_z\rangle=0$
 then this implies $\langle d^{\dagger} d\rangle=\frac{1}{2}$. In average, the second term is zero in the presence of a small magnetic field. A similar physical conclusion for the first term then should be reached 
 owing to the symmetry between the two apparatus or impurities. In fact, we could have equally absorbed the effect of commutation relations between the two spins
 as a Jordan-Wigner string $e^{i \pi f^{\dagger} f}$ on the first spin. The first term would be modified as $-h(a+b)(1-2f^{\dagger} f)$ with $\langle f^{\dagger} f\rangle=\frac{1}{2}$ such that $\langle \tau_z\rangle=0$.
 Therefore, in average a small magnetic field will not modify the ground state and the Majorana fermions $a$ and $m$ remain free. 
 
If we develop the free energy to second order in perturbation theory, we have a correction in $\sim \int_0^{\frac{1}{\Delta}} d\tau\delta H^2$. We have e.g. the identification 
$\frac{1}{2}h_0^2 \frac{1}{\Delta} e^{-2i\omega t} b(1-2d^{\dagger} d)m=-\frac{1}{2}h_0^2 \frac{1}{\Delta}e^{-2i\omega t}i a m$. Since the Hamiltonian is hermitian this allows for an oscillatory term in time of the form
\begin{equation}
\delta H' = -h_{eff} \cos(2\omega t) i a m.
\end{equation}
This would allow in time to adjust (flip) the parity state of the operator $i am$, associated to the zero-energy Majorana fermions $a$ and $m$, from $+$ to $-$ in time when $\cos(2\omega t) = \mp$ respectively. Flipping the sign of  $\cos(2\omega t)$ is also similar as
inverting the position of $a$ and $m$ while preserving $\langle S_z\rangle=\langle \tau_z\rangle=0$ i.e. $\langle d^{\dagger} d\rangle=\langle f^{\dagger} f\rangle = \frac{1}{2}$. 

The operator $\langle i am \rangle$ can be measured from the spin correlation function $\langle S_y \tau_x\rangle$. A similar form of $\delta H'$ may in principle be reached with a rotating magnetic field i.e. inducing a form of perturbation such as $h_0 e^{i\omega t}(S_x + i S_y)+h.c.$ on one impurity and similarly for the other impurity. This can also be realized with circularly polarized light \cite{fractionalinfo}.

Similarly as  systems of two spins \cite{fractionalinfo} or two quantum dots \cite{Flensberg,Delft1,Delft}, we can then form a delocalized spin-1/2 $\boldsymbol\sigma$, qubit (fermion), such that $F=\frac{1}{\sqrt{2}}(a+im)=\sigma^-$ and 
$F^{\dagger} F-\frac{1}{2}= i am=\sigma_z$.

\bibliographystyle{crunsrt}
\bibliography{samplebib}

\end{document}